\documentclass[12pt]{article}
\pdfoutput=1

\usepackage{amsmath,amssymb}
\usepackage{graphicx}

\makeatletter

\renewcommand\section{\@startsection{section}{1}{\z@}
                                   {-3.5ex \@plus -1ex \@minus -.2ex}
                                   {2.3ex \@plus .2ex}
                                   {\normalfont\large\bfseries}}
\renewcommand\subsection{\@startsection{subsection}{2}{\z@}
                                   {-3.25ex\@plus -1ex \@minus -.2ex}
                                   {1.5ex \@plus .2ex}
                                   {\normalfont\normalsize\bfseries}}
\renewcommand\subsubsection{\@startsection{subsubsection}{3}{\z@}
                                   {-3.25ex\@plus -1ex \@minus -.2ex}
                                   {1.5ex \@plus .2ex}
                                   {\normalfont\normalsize\bfseries}}
\renewcommand\paragraph{\@startsection{paragraph}{4}{\z@}
                                   {3.25ex \@plus1ex \@minus.2ex}
                                   {-1em}
                                   {\normalfont\normalsize\bfseries}}

\makeatother

\newcommand{\beq}{\begin{equation}}
\newcommand{\eeq}{\end{equation}}
\newcommand{\bea}{\begin{eqnarray}}
\newcommand{\eea}{\end{eqnarray}}

\newcommand{\SU}{{\rm SU}}
\newcommand{\SO}{{\rm SO}}

\newcommand{\Spin}{\rm Spin}

\newcommand{\R}{\mathbb R}

\newcommand{\id}{\hbox{1\kern-.27em l}}

\newcommand{\ad}{{\rm ad}}

\newcommand{\cO}{{\cal O}}

\begin{document}

\pagestyle{empty}

\begin{center}

\vspace*{30mm}
{\Large Tunneling solutions in topological field theory on $\R \times S^3 \times I$}

\vspace*{30mm}
{\large Louise Anderson and M{\aa}ns Henningson}

\vspace*{5mm}
Department of Fundamental Physics\\
Chalmers University of Technology\\
S-412 96 G\"oteborg, Sweden\\[3mm]
{\tt louise.anderson@chalmers.se, mans@chalmers.se}

\vspace*{30mm}{\bf Abstract:}
\end{center}
We consider a topologically twisted version, recently introduced by Witten, of five-dimensional maximally supersymmetric Yang-Mills theory on a five-manifold of the form $M_5 = \R \times W_3 \times I$. If the length of the interval $I$ is sufficiently large, the supersymmetric localization equations admit pairs of static solutions (with the factor $\R$ interpreted as Euclidean time). However, these solutions disappear for a sufficiently short $I$, so by the topological invariance of the theory, they must be connected by an interpolating dynamic instanton solution. We study this for the case that $W_3$ is a three-sphere $S^3$ with the standard metric by making a spherically symmetric Ansatz for all fields. The solution is given as a power series in a parameter related to the length of $I$, and we give explicit expressions for the first non-trivial terms.

\newpage \pagestyle{plain}

\section{Introduction}

On a five-manifold of the form
\beq
M_5 = M_4 \times I ,
\eeq
where $M_4$ is an oriented Riemannian four-manifold and $I$ is an interval, five-dimensional maximally supersymmetric Yang-Mills theory admits several inequivalent topological twistings. One of these was recently introduced by Witten \cite{Witten2011} in an attempt to find a Yang-Mills interpretation of the Khovanov homology of knots \cite{Khovanov}. This twisting corresponds to a homomorphism from the $\Spin (4)$ holonomy group of $M_5$ to the $\Spin (5)$ $R$-symmetry group of the Yang-Mills theory under which the spinor representation ${\bf 4}$ of $\Spin (5)$ decomposes as a direct sum ${\bf 2} + {\bf 2}$ of two chiral spinor representations of $\Spin (4)$. After such a twisting, the bosonic degrees of freedom can be described by fields on $M_4$ that in addition depend on a linear coordinate $y$ on $I$. These are the gauge connection $A$ with field strength $F$, a two-form $B$ which is self-dual with respect to the orientation and Riemannian structure of $M_4$, and a complex zero-form $\sigma$. The fields $F$, $B$, and $\sigma$ take their values in the vector bundle $\ad (E)$ associated to the gauge bundle $E$ via the adjoint representation of the gauge group $G$.

The path integral of this topological field theory localizes on supersymmetric field configurations that obey the elliptic set of equations \cite{Witten2011}
\bea \label{d=5_equations}
F^+ - \frac{1}{4} B \times B - \frac{1}{2} D_y B & = & 0 \cr
F_y + * D B & = & 0 .
\eea
Here $*$ denotes the Hodge duality operator constructed from the orientation and Riemannian structure on $M_4$, $\times$ is an antisymmetric `cross-product' on the rank three vector bundle $\Omega^{2, +} (M_4)$ of self-dual two forms on $M_4$, $D$ is the gauge covariant exterior derivative on $M_4$, and $D_y$ is the covariant derivative with respect to the coordinate $y$ on $I$. Furthermore, we have decomposed the field strength $F$ (which is a two-form on $M_5$) as $F = F_y + F^+ + F^-$, where the three terms are a one-form, a self-dual two-form and an anti-self-dual two-form on $M_4$ respectively. In addition to (\ref{d=5_equations}), supersymmetric field configurations obey some further equations that typically are equivalent to demanding that $\sigma$ vanish identically.

At the endpoints of $I$, the above bulk equations must be supplemented by suitable elliptic boundary conditions. (See the appendix of \cite{Witten2010} for an accessible introduction to this concept.) Those that we will be using can be described as follows in terms of the Riemannian geometry of $M_4$ \cite{Witten2011}: Suppose first that the gauge group $G$ is isomorphic to $\SO (3)$. As we approach a boundary component of $M_5$, the gauge connection $A$ tends to the Riemannian connection on the rank three vector bundle $\Omega^{2, +} (M_4)$. The self-dual two-form $B$ has a simple pole at the boundary, the residue of which is given by the self-dual part of the second exterior power of the vierbein on $M_4$. (See \cite{Henningson2011} for a detailed discussion of a similar set of boundary conditions in four dimensions.) For a general gauge group $G$, the boundary behaviors of $A$ and $B$ are obtained in the same way, except that we have to identify $\SO (3)$ with a principally embedded subgroup of $G$ \cite{Kostant}.

In this paper, we will only be concerned with the special case when
\beq
M_4 = \R \times W_3 ,
\eeq
where the first factor can be interpreted as an Euclidean time direction and $W_3$ is a Riemannian three-manifold. We can then identify $B$ with a one-form $\phi$ on $W_3$ with values in $\ad (E)$. It is convenient to work in temporal gauge in which the time-component of the gauge field vanishes. We also single out the component of the gauge field in the direction of $I$ and denote this with  $\chi$, (which can be regarded as a zero-form on $W_3$), so that $A$ and $F$ henceforth denote the components of the gauge field and its field strength in the directions of $W_3$ only. With $*$ and $D$ henceforth denoting the Hodge duality operator and the covariant exterior derivative respectively on $W_3$, the equations (\ref{d=5_equations}) then amount to
\bea \label{dynamic_equations}
\frac{\partial A}{\partial t} & = & - * (F - \phi \wedge \phi) + D_y \phi \cr
\frac{\partial \phi}{\partial t} & = & * D \phi - F_y \cr
\frac{\partial \chi}{\partial t} & = & - * D (* \phi) .
\eea

Static, i.e. $t$-independent, solutions to (\ref{dynamic_equations}) are of particular importance. Imposing the additional gauge condition that the component $\chi$ of the gauge field in the direction of $I$ vanish\footnote{This is possible by performing a gauge transformation with a time-independent parameter.}, such solutions amount to configurations on $W_3 \times I$ obeying the equations
\bea \label{static_equations}
\frac{\partial \phi}{\partial y} & = & * (F - \phi \wedge \phi) \cr
\frac{\partial A}{\partial y} & = & * D \phi \cr
* D (* \phi) & = & 0 .
\eea
This can be interpreted as describing a perturbative ground state of the quantum theory on the spatial manifold $W_3 \times I$. However, the energy of a bose-fermi pair of such states, i.e. a pair of solutions to the static equations (\ref{static_equations}), may become lifted by a non-perturbative tunneling effect represented by an interpolating solution to the dynamic equations (\ref{dynamic_equations}) \cite{Witten2011}. This phenomenon is well-known from applications of Morse theory to supersymmetric quantum mechanics \cite{Witten1982} or, in an infinite dimensional setting, of Floer homology \cite{Floer} to topological quantum field theory \cite{Witten1988}. As usual, while the space of perturbative ground states may depend on the precise geometry of $W_3 \times I$, the space of exact ground states is a topological invariant. In particular, with a fixed Riemannian structure on $W_3$ and a sufficiently short interval $I$, we would not expect the static equations (\ref{static_equations}) to have any solutions at all.\footnote{As described above, while the gauge field $A$ should tend to the same finite limit at both boundary components, the one-form $\phi$ should have poles with opposite non-zero residues. This behavior does not seem to be compatible with an arbitrarily short distance between the components.} So any solutions to (\ref{static_equations}) that may occur under particular circumstances should be pairwise connected by interpolating solutions to (\ref{dynamic_equations}).

The aim of this note is to give an explicit example of such a tunneling solution. To this end, we will in the next section take $W_3 = S^3$ with the standard metric, and make a spherically symmetric Ansatz for all fields. In section three, we will in this setting discuss the existence of a pair of solutions to the static equations (\ref{static_equations}) that were found in \cite{Henningson2011} provided that the interval $I$ has a length $\Delta y > 2 \pi$. For $\Delta y = 2 \pi$ the two solutions coalesce and they disappear for $\Delta y < 2 \pi$. These solutions can be approximated by a power series in a parameter $\varepsilon$ related to the interval length so that $\varepsilon = 0$ for $\Delta y = 2 \pi$ and $\varepsilon \rightarrow \pm 1$ as $\Delta y \rightarrow \infty$. In section four, we will then discuss the corresponding approximate tunneling solution to (\ref{dynamic_equations}), also as a power series in $\varepsilon$. We construct the first few terms explicitly and indicate how to proceed to arbitrary orders, although a rigorous proof of existence of the solution is still lacking.

While the present work is clearly concerned with a rather particular situation, it should be possible to generalize it in many different directions. In particular, one would hope that our results may be helpful for further developing the Yang-Mills interpretation \cite{GaiottoWitten2009} of the Jones polynomial \cite{Jones} and Khovanov homology \cite{Khovanov} of knot theory.

\section{Spherical symmetry}
To be able to find explicit solutions to the equations, we will henceforth specialize to the case that
\beq
W_3 = S^3
\eeq
with the standard metric. It is convenient to use the property that $S^3$ is isomorphic to the group manifold $\SU (2)$ and choose the dreibein $e$ as the Lie algebra valued Mauer-Cartan form obeying
\bea
d e & = & -e \wedge e \cr
& = & - * e .
\eea
The corresponding spin connection is then $\omega = \frac{1}{2} e$.

The most general spherically symmetric Ansatz for the gauge field $A$, the one-form $\phi$ and the zero-form $\chi$ (the component of the gauge field along $I$) is
\bea
A & = & \frac{1}{2}(1 + u) e \cr
\phi & = & s e \cr
\chi & = & 0 ,
\eea
where $u$ and $s$ are functions of time $t$ and the linear coordinate $y$ on $I$ only. The field $\chi$, which in general is a zero-form on $W_3$, is put to zero identically, since there is no non-vanishing possible spherically symmetric Ansatz for it. The last equation in (\ref{dynamic_equations}) is then identically satisfied while the first two read
\bea \label{dynamic_symmetric}
\frac{\partial u}{\partial t} & = & 2 \frac{\partial s}{\partial y} - \frac{1}{2} u^2 + 2 s^2 + \frac{1}{2} \cr
\frac{\partial s}{\partial t} & = & - \frac{1}{2} \frac{\partial u}{\partial y} + s u .
\eea
The boundary conditions are that $u$ vanishes and that $s$ has simple poles with residues $\pm 1$ at the endpoints of the interval $I$.

The static equations (\ref{static_equations}) are of course obtained by requiring $u = \tilde{u}$ and $s = \tilde{s}$ to depend on $y$ only and read
\bea \label{static_symmetric}
\frac{d \tilde{u}}{d y} & = & 2 \tilde{s} \tilde{u} \cr
\frac{d \tilde{s}}{d y} & = & \frac{1}{4} \tilde{u}^2 - \tilde{s}^2 - \frac{1}{4} .
\eea

\section{The static solutions}

\begin{figure}
\centering
\includegraphics[width=0.7\textwidth]{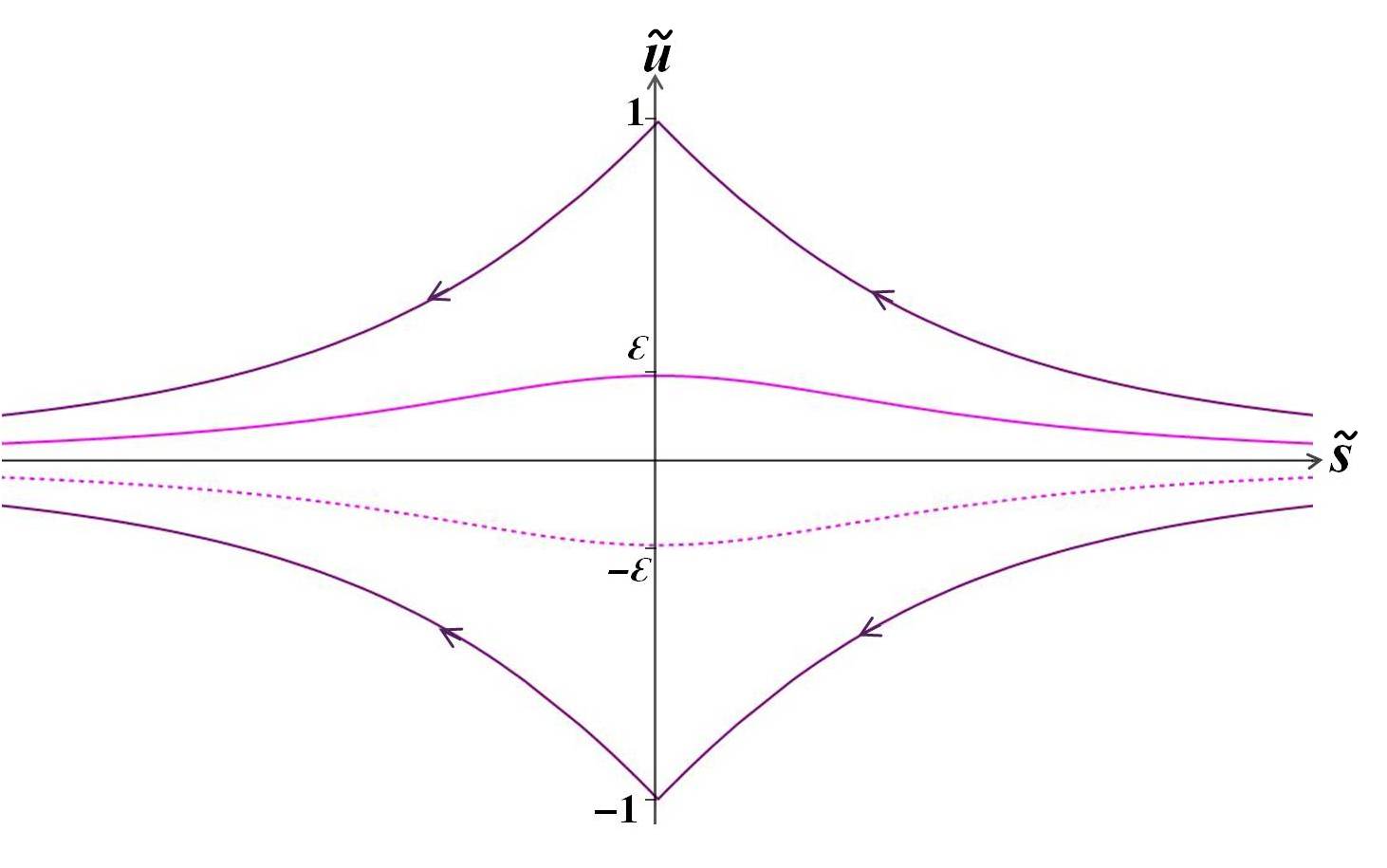}
\caption{The flow in the $\tilde{s}-\tilde{u}$-plane, where arrows denote direction of increasing y. Only solutions in the interior of the space defined by the solutions flowing to or from the critical points $\tilde{s}=0$, $\tilde{u}=\pm1$ will satisfy our boundary conditions. They appear in pairs related by reflection in the $\tilde{s}$-axis.}
\label{fig:SU-plane}
\end{figure}

The static equations (\ref{static_symmetric}) can be visualized as a flow in the $\tilde{s} \tilde{u}$-plane, as is seen in figure \ref{fig:SU-plane} with arrows indicating the direction of increasing $y$.
We have two critical points, namely $\tilde{s}=0, \tilde{u}=\pm 1$, which correspond to the trivial configurations $\phi = 0, A=0$ and $\phi=0, A=e$ respectively. (The two critical points are in fact related to each other by a `large' gauge transformation and are thus physically equivalent \cite{Henningson2011}.) In the interior of the region bordered by the solutions flowing to and from these critical points, we have solutions that satisfy our boundary conditions. All solutions start at $\tilde{s} = \infty$, $\tilde{u} = 0$ and end at $\tilde{s} = - \infty$, $\tilde{u} = 0$ corresponding to the two boundary components of $W_3 \times I$. The divergences of $\tilde{s}$ and the zeros of $\tilde{u}$ reflect the Nahm-poles of $\phi$ and the limits $\omega = \frac{1}{2} e$ of $A$ as we approach the boundaries.

We fix the symmetry under translations in $y$ by demanding that $\tilde{s}$ be an even function of $y$, from which it follows that $\tilde{u}$ is an odd function of $y$. The solutions can then be parametrized by a single integration constant which we take to be
\beq
\varepsilon = \tilde{u} (0)
\eeq
with $-1 < \varepsilon < 1$. Clearly, the solutions appear in pairs related by reflection in the $s$-axis, i.e. $s \rightarrow s$ and $u \rightarrow - u$ so that $\varepsilon \rightarrow - \varepsilon$. A typical pair of such solutions has been indicated in figure \ref{fig:SU-plane}.

The parameter $\varepsilon$ is related to the interval length $\Delta y$, which (provided that the functions $\tilde{u}$ and $\tilde{s}$ are known) can be computed as
\bea
\Delta y & = & \int d y \cr
& = & \int_\infty^{-\infty} d \tilde{s} \Big/ \frac{d \tilde{s}}{d y} \cr
& = & \int_\infty^{-\infty} d \tilde{s} \Big/ \left(\frac{1}{4} \tilde{u}^2 - \tilde{s}^2 - \frac{1}{4} \right) .
\eea
The interval length is a monotonously increasing function of $\varepsilon^2$ with $\Delta y = 2 \pi$ for $\varepsilon = 0$ and $\Delta y \rightarrow \infty$ as $\varepsilon \rightarrow \pm 1$. 

This limit corresponds to solutions that approach one of the critical points as $y \rightarrow \infty$ (or $y \rightarrow - \infty$) and get stuck there;  this means that the finite interval $I$ gets replaced by the half-line $\mathbb{R}^+$. This is actually the case that is most relevant for the study of knot theory \cite{Witten2011}. It is also the only case in which we will have a non-zero instanton number. This can be seen as follows:
The instanton density is locally a total derivative of the Chern-Simons invariants. The instanton number is thus given by considering the difference between them at the endpoints of the interval $I$. In cases where $I$ is finite, this will be trivially zero since the Chern-Simons invariants at the endpoints are equal. In the case where we have the entire half-line, our solutions interpolates between a trivial gauge field (at $y=\infty$) and a gauge field given by the spin connection on $S^3$ (at $y=0$). This results in an instanton number of $\pm 1$ for each of the critical points respectively.

For non-zero $\varepsilon$, it does not seem possible to find closed expressions for $\tilde{s}$ and $\tilde{u}$, but if we express them as odd and even power series respectively in $\varepsilon$
\bea
\tilde{s} & = & \tilde{s}_0 + \varepsilon^2 \tilde{s}_2 + \ldots \cr
\tilde{u} & = & \varepsilon \tilde{u}_1 + \varepsilon^3 \tilde{u}_3 + \ldots ,
\eea
the equations (\ref{static_symmetric}) amount to an infinite system of differential equations for the coefficients:
\bea
\frac{\partial \tilde{s}_0}{\partial y} & = & - \tilde{s}_0^2 - \frac{1}{4} \cr
\frac{\partial \tilde{u}_1}{\partial y} & = & 2 \tilde{s}_0 \tilde{u}_1 \cr
\frac{\partial \tilde{s}_2}{\partial y} & = & \frac{1}{4} \tilde{u}_1^2 - 2 \tilde{s}_0 \tilde{s}_2 \cr
& \ldots &
\eea
It is straightforward to solve these recursively: Together with the condition that $\tilde{s}$ be an odd function of $y$, the first equation yields
\beq
\tilde{s}_0 = - \frac{1}{2} \tan \frac{y}{2} .
\eeq
When this is inserted in the second equation, the solution is
\beq
\tilde{u}_1 = \cos^2 \frac{y}{2} ,
\eeq
where the integration constant is fixed by the requirement that $\tilde{u}_1 (0) = 1$. The unique odd solution to the third equation is then
\beq \label{tildes2}
\tilde{s}_2 =  \frac{5 y}{64} \cos^{-2} \frac{y}{2} + \sin \frac{y}{2} \left( \frac{5}{32} \cos^{-1} \frac{y}{2} + \frac{5}{48} \cos \frac{y}{2} + \frac{1}{12} \cos^3 \frac{y}{2} \right) .
\eeq
This procedure can obviously be generalized to arbitrary orders in $\varepsilon$. We have also computed $\tilde{u}_3$ and $\tilde{s}_4$ explicitly, but the expressions are rather unilluminating and will be omitted here.

A general feature is that $\tilde{s}_{2 n}$ will have poles of order $n + 1$ and $\tilde{u}_{2 n + 1}$ will have poles of order $n - 2$ at $y = \pm \pi$. (Poles of negative order are of course interpreted as zeros, so that $\tilde{u}_1$ has second order zeros as seen above, $\tilde{u}_3$ has first order zeros, and $\tilde{u}_5$ is regular and non-vanishing at $y = \pm \pi$.) This is related to the fact that $\tilde{u}$ and $\tilde{s}$ should vanish and have first order poles of unit residues respectively at the ends of the interval $I$. The length $\Delta y$ of $I$ depends on $\varepsilon$ in a way that dictates the singularity structure at $y = \pm \pi$ of the coefficients $\tilde{s}_{2 n}$ and $\tilde{u}_{2 n + 1}$. To second order in $\varepsilon$ we have
\beq
\Delta y = 2 \pi + \varepsilon^2 \frac{5 \pi}{8} + \cO (\varepsilon^4) .
\eeq

\section{The tunneling solution}
As described in the introduction, we expect the pair of solutions $\pm \tilde{u}$ and $\tilde{s}$ to the static equations (\ref{static_symmetric}) obtained for some fixed interval length $\Delta y > 2 \pi$, to be connected by an interpolating tunneling solution $u (y, \tau)$ and $s (y, \tau)$. The boundary conditions in the time direction are thus that
\bea
u (y, \tau) \rightarrow \pm \tilde{u} (y) \cr
s (y, \tau) \rightarrow \tilde{s} (y)
\eea
as $\tau \rightarrow \pm \infty$. At the endpoints of the spatial interval $I$ for finite $\tau$, we require as before that $u$ vanishes and $s$ has simple poles with residues $\pm 1$. Here we have introduced a rescaled time variable $\tau$ as
\beq
\tau = \varepsilon t
\eeq
in terms of which the equations (\ref{dynamic_symmetric}) read
\bea \label{dynamic_tau}
\varepsilon \frac{\partial u}{\partial \tau} & = & 2 \frac{\partial s}{\partial y} - \frac{1}{2} u^2 + 2 s^2 + \frac{1}{2} \cr
\varepsilon \frac{\partial s}{\partial \tau} & = & - \frac{1}{2} \frac{\partial u}{\partial y} + s u .
\eea
In this way, it is consistent that $u$ and $s$ for fixed $y$ and $\tau$ be given by odd and even power series in $\varepsilon$ respectively:
\bea
s & = & s_0 + \varepsilon^2 s_2 + \ldots \cr
u & = & \varepsilon_1 u_1 + \varepsilon^3 u_3 + \ldots .
\eea
The equations (\ref{dynamic_tau}) then amount to an infinite system of differential equations:
\bea
0 & = & 2 \frac{\partial s_0}{\partial y} + 2 s_0^2 + \frac{1}{2} \cr
\frac{\partial s_0}{\partial \tau} & = & - \frac{1}{2} \frac{\partial u_1}{\partial y} + s_0 u_1 \cr
\frac{\partial u_1}{\partial \tau} & = & 2 \frac{\partial s_2}{\partial y} - \frac{1}{2} u_1^2 + 4 s_0 s_2 \cr
& \ldots &
\eea
Together with the condition that $s$ be an odd function of $y$, the first equation gives
\beq
s_0 = - \frac{1}{2} \tan \frac{y}{2} .
\eeq
When this is inserted in the second equation, we get a differential equation for $u_1$ as a function of $y$ (for fixed $\tau$) with the general solution
\beq
u_1 = c_1 (\tau) \cos^2 \frac{y}{2} .
\eeq
To determine the $\tau$-dependent integration constant $c_1$, we need to consider the third equation, which now reads
\beq
\frac{\partial c_1}{\partial \tau} \cos^2 \frac{y}{2} = 2 \frac{\partial s_2}{\partial y} - \frac{1}{2} c_1^2 \cos^4 \frac{y}{2} - 2 s_2 \tan \frac{y}{2} .
\eeq
If this is regarded as an ordinary differential equation for $s_2$ as a function of $y$, the unique odd solution for an arbitrary function $c_1$ is
\bea \label{s2_preliminary}
s_2 & = & \frac{\partial c_1}{\partial \tau} \cos^{-2} \frac{y}{2} \left( \frac{3}{16} y + \sin \frac{y}{2} \left( \frac{3}{8} \cos \frac{y}{2} + \frac{1}{4} \cos^3 \frac{y}{2} \right) \right) \cr
& & + c_1^2 \cos^{-2} \frac{y}{2} \left( \frac{5}{64} y + \sin \frac{y}{2} \left( \frac{5}{32} \cos \frac{y}{2} + \frac{5}{48} \cos^3 \frac{y}{2} + \frac{1}{12} \cos^5 \frac{y}{2} \right) \right) .
\eea
The first and second term on each line gives rise to second order and first order poles at $y = \pm \pi$ respectively. These must agree with the poles in $\tilde{s}_2$ found in the previous section, which gives the condition
\beq
\frac{5}{64} = \frac{3}{16} \frac{\partial c_1}{\partial \tau} + \frac{5}{64} c_1^2 .
\eeq
Up to an inessential translation of the time-variable $\tau$, which we fix by requiring $c_1$ to be an odd function of $\tau$, the unique solution to this differential equation is $c_1 = \tanh \frac{5 \tau}{12}$ so that
\beq
u_1 = \tanh \frac{5 \tau}{12} \cos^2 \frac{y}{2} .
\eeq
As required for an interpolating solution, $u_1 \rightarrow \pm \tilde{u}_1 = \pm \cos^2 \frac{y}{2}$ as $\tau \rightarrow \pm \infty$.

The corresponding solution (\ref{s2_preliminary}) is best presented in the form
\beq
s_2 = \tilde{s}_2 + d_2 ,
\eeq
where the static term $\tilde{s}_2$ is given in (\ref{tildes2}) and the `dynamic' term $d_2$ is given by the surprisingly simple result
\beq
d_2 = - \frac{1}{12} \cosh^{-2} \frac{5 \tau}{12} \sin \frac{y}{2} \cos^3 \frac{y}{2} .
\eeq
We note that not only is $d_2$ regular at $y = \pm \pi$ as required by the boundary conditions, but in fact it has third order zeros there. Moreover, as required for an interpolating solution, $d_2 \rightarrow 0$ as $\tau \rightarrow \pm \infty$.

The generalization of this procedure to arbitrary orders in $\varepsilon$ is not as obvious as in the static case, but we believe that it can be carried out. A rigorous proof of this would be quite interesting. An outline is as follows: When the functions $s_0, u_1, \ldots, s_{2 n}$ have been determined, the terms of order $2 n + 1$ and $2 n + 2$ in $\varepsilon$ respectively in equations (\ref{dynamic_tau}) can be written as
\bea
\frac{\partial s_{2 n}}{\partial \tau} & = & - \frac{1}{2} \frac{\partial u_{2 n + 1}}{\partial y} - \frac{1}{2} u_{2 n + 1} \tan \frac{y}{2} + \Bigl[(s u)_{2 n + 1} - s_0 u_{2 n +1} \Bigr] \cr
\frac{\partial u_{2 n + 1}}{\partial \tau} & = & 2 \frac{\partial s_{2 n + 2}}{\partial y} - u_{2 n + 1} \tanh \frac{5 \tau}{12} \cos^2 \frac{y}{2} - 2 s_{2 n + 2} \tan \frac{y}{2} \cr
& & + \left[ - \frac{1}{2} (u^2)_{2 n + 2} + u_1 u_{2 n + 1} + 2 (s^2)_{2 n + 2} - 4 s_0 s_{2 n + 2} \right] ,
\eea
where we have used the expressions for $s_0$ and $u_1$ given above. It is important to note that the expressions in square brackets are given in terms  of the previously determined functions $s_0, u_1, \ldots, s_{2 n}$. The general solution to the first equation, regarded as an ordinary differential equation for $u_{2 n + 1}$ as a function of $y$ for fixed $\tau$, is
\beq
u_{2 n + 1} = 2 I_{2 n + 1} \cos^2 \frac{y}{2}  ,
\eeq
where
\beq
I_{2 n + 1} = \int d y \cos^{-2} \frac{y}{2}\Bigl[(s u)_{2 n + 1} - s_0 u_{2 n +1} - \frac{\partial s_{2 n}}{\partial \tau} \Bigr] .
\eeq
Clearly, the indefinite integral $I_{2 n +1}$ is only determined up to an arbitrary ($\tau$-dependent) integration constant. We insert this result in the second equation, which then reads
\bea
2 \frac{\partial I_{2 n + 1}}{\partial \tau} \cos^2 \frac{y}{2} & = & 2 \frac{\partial s_{2 n + 2}}{\partial y} - 2 I_{2 n + 1} (\tau) \tanh \frac{5 \tau}{12} \cos^4 \frac{y}{2} - 2 s_{2 n + 2} \tan \frac{y}{2} \cr
& & + \left[ - \frac{1}{2} (u^2)_{2 n + 2} + u_1 u_{2 n + 1} + 2 (s^2)_{2 n + 2} - 4 s_0 s_{2 n + 2} \right] .
\eea
The solution to this equation, regarded as an ordinary differential equation for $s_{2 n + 2}$ as a function of $y$ for fixed $\tau$, is
\pagebreak
\bea
s_{2 n + 2} & = & \cos^{-2} \frac{y}{2} \int d y \left\{ \frac{\partial I_{2 n + 1}}{\partial \tau} \cos^4 \frac{y}{2} + I_{2 n+ 1} \tanh \frac{5 \tau}{12} \cos^6 \frac{y}{2} \right. \cr
& & - \left. \frac{1}{2} \cos^2 \frac{y}{2} \left[ - \frac{1}{2} (u^2)_{2 n + 2} + u_1 u_{2 n + 1} + 2 (s^2)_{2 n + 2} - 4 s_0 s_{2 n + 2} \right] \right\} .
\eea
Imposing that $s_{2 n + 2}$ should be an odd function of $y$ determines the integration constant uniquely. The function $s_{2 n + 2}$ of course depends on $\tau$, and has poles at $y = \pm \pi$ of order up to $n + 2$. But hopefully, it is of the form
\bea
s_{2 n + 2} & = & A_{2n + 2} (\tau) \frac{\partial}{\partial y} \left( y \tan \frac{y}{2} \right) + \mathrm{static \; terms} + \mathrm{regular \; terms}
\eea
for some arbitrary (even) function $A_{2 n + 2}$ of $\tau$, i.e. the $\tau$-dependent part of $s_{2 n + 2}$ should have no poles beyond second order, and the first and second order poles should be given by the $y$-derivative of a first order pole. We believe that it should be possible to prove this property by a careful study of the structure of the singularities and time-dependence of the coefficients of $u$ and $s$. The unwanted $\tau$-dependent singular terms can now be removed by an appropriate choice of the integration constant in $I_{2 n + 1}$. Indeed, shifting
\beq
I_{2n + 1} \rightarrow I_{2 n + 1} + C_{2 n + 1} (\tau)
\eeq
gives shifts
\bea
u_{2 n + 1} & \rightarrow & u_{2 n + 1} + 2 C_{2 n + 1} \cos^2 \frac{y}{2} \cr
s_{2 n + 2} & \rightarrow & s_{2 n + 2} + \left( \frac{d C_{2 n + 1}}{d \tau} + \frac{5}{6} C_{2 n +1} \tanh \frac{5 \tau}{12} \right) \frac{3}{4} \frac{\partial}{\partial y} \left( y \tan \frac{y}{2} \right) \cr
& & + \left( \frac{d C_{2 n + 1}}{d \tau} + \frac{5}{6} C_{2 n +1} \tanh \frac{5 \tau}{12} \right) \frac{1}{2} \sin \frac{y}{2} \cos \frac{y}{2} \cr
& & + \frac{1}{3} C_{2 n + 1} \tanh \frac{5 \tau}{12} \sin \frac{y}{2} \cos^3 \frac{y}{2} .
\eea
The function $C_{2 n + 1}$ should thus be chosen to satisfy the ordinary differential equation
\beq
\frac{d C_{2 n + 1}}{d \tau} + \frac{5}{6} C_{2 n +1} \tanh \frac{5 \tau}{12} = - \frac{4}{3} A_{2 n + 2} ,
\eeq
the solution of which is
\beq
C_{2 n + 1} = - \frac{4}{3} \cosh^{-2} \frac{5 \tau}{12} \int d \tau A_{2 n + 2} \cosh^2 \frac{5 \tau}{12}
\eeq
with the integration constant fixed by the requirement that $I_{2 n + 1}$ be an odd function of $\tau$. After this shift, the new function $s_{2 n + 2}$ will only have time independent singular terms, which of course should agree with the singularities of the static solution $\tilde{s}_{2 n + 2}$.

We have computed the functions $u_3$ and $s_4$ explicitly along these lines, but the resulting expressions are even less illuminating than in the static case and will be omitted here.

\vspace*{5mm}
This research was supported by grants from the G\"oran Gustafsson foundation and the Swedish Research Council.

\bibliographystyle{naturemag}

\bibliography{references}

\end{document}